# Scalable and Tunable In-Plane Ge/Si(001) Nanowires Grown by Molecular Beam Epitaxy

Jian-Huan Wang[1,2,3], Ming Ming[2,4], Ding-Ming Huang[1,2,3], Jie-Yin Zhang[2,3,4], Yi Luo[1,3], Bin-Xiao Fu[2,4], Yi-Xin Chu[2], Yuan Yao[2], Hongqi Xu[1,3,*] and Jian-Jun Zhang[2,4,5,*]

[1]*Beijing Academy of Quantum Information Sciences, Beijing 100193, China*
[2]*Beijing National Laboratory for Condensed Matter Physics and Institute of Physics, Chinese Academy of Sciences, Beijing 100190, China*
[3]*Beijing Key Laboratory of Quantum Devices and School of Electronics, Peking University, Beijing 100871, China*
[4]*Songshan Lake Materials Laboratory, Guangdong 523808, China*
[5]*Hefei National Laboratory, Hefei 230088, China*

(Dated: June 24, 2025)

**Abstract**: Germanium nanostructures offer significant potential in developing advanced integrated circuit and disruptive quantum technologies, yet achieving both scalability and high carrier mobility remains a challenge in materials science. Here, we report an original low-temperature epitaxial method for growth of site-controlled in-plane germanium nanowires with high hole mobility by molecular beam epitaxy. By reducing the growth temperature, we effectively suppress Si-Ge interdiffusion, ensuring pure germanium composition within the nanowires while preserving their high crystalline quality. The method employs pre-patterned ridges on strain-relaxed $Si_{0.75}Ge_{0.25}$/Si(001) substrates as tailored templates, enabling control over the position, length, spacing and cross-sectional shape of the nanowires. Electrical measurements of field-effect devices made from as-grown germanium nanowires show that the nanowires are of hole conduction with mobility exceeding 7000 $cm^2$/Vs at 2-20 K. The method paves a way for fabrication of scalable germanium nanowire networks, providing a reliable platform for the developments of high-performance nanoelectronics and multi-qubit chips.

With advancements in microelectronics technology, especially in low-temperature processing, high-$\kappa$ dielectric materials, and heterogeneous integration of Ge on Si, germanium has re-emerged as a vital transistor channel material with renewed vigor [1-4]. In response to the latest transistor architectures proposed for advanced nodes, one-dimensional germanium (1D Ge), inheriting high mobility from bulk materials, naturally emerges as a promising channel material for gate-all-around transistors [5-8]. Concurrently, with the rapid development of quantum technology, 1D Ge also holds potential as a novel material for constructing hybrid quantum circuits. To date, spin qubits [9-11], known for their small footprints, and superconducting qubits [12,13], which enable high-fidelity readout and control, have been realized in this material platform. Furthermore, theoretical predictions [14,15] suggest the potential for using 1D Ge in constructing topological qubits [16-19] that are immune to environmental noise.

A scalable, high-quality 1D Ge material platform is essential for achieving high-performance transistors and enabling efficient multi-qubit quantum computing chips. Top-down technique has been used for conventional transistor fabrication due to its advantages in scalability and yield. However, top-down etching introduces surface defects [20], which deteriorate the material properties. In contrast, Ge nanowires (NWs) obtained by bottom-up method typically offer a higher material quality. For example, a high hole mobility of 4400 cm$^2$/Vs at 4 K has been achieved in VLS-grown Ge/Si core/shell NWs orientated in a <110> crystallographic direction [21]. Self-assembled Ge NWs with high hole mobility are found to exhibit strong spin-orbit coupling [22-24], enables substantial improvements in qubit manipulation speed [10,11,25]. Experimentally, Rabi frequencies exceeding 1.2 GHz have been demonstrated [25]. On the other hand, theoretical studies [26,27] have predicted that optimizing NW orientation and geometry can suppress the influence of charge noise and even of nuclear spin noise, thereby extending coherence time in qubits made from Ge NWs. Experimentally, an evidence for charge noise suppression in a Ge hole double quantum dot has been reported [28]. Regarding the development of qubit arrays, research based on Ge NWs lags behind that of the planar Ge/SiGe heterostructures. Notably, most recently reported array architectures [29,30] in the planar heterostructures are transferable to NW networks with high carrier mobility. Furthermore, the inherent 2D confinement in NWs enables potential reduction in number of gate electrodes in qubit definition.

Notwithstanding considerable progress in NW growth made so far, it remains a formidable challenge to obtain scalable Ge NW arrays with high carrier mobility. While the extensively studied VLS-grown NWs exhibit high carrier mobility, as previously described, these NWs are mainly in out-of-plane, free-standing forms [6,21,31]. Recently, in-plane Ge NWs [32-34] on Si(001) surfaces have been achieved with selective-area growth technique. This approach ingeniously leverages the distinct sticking coefficients of germane on Si and SiO$_2$, enabling site-selective growth of Ge NWs on a nanoslit-patterned SiO$_2$/Si(001) substrate. However, a relatively high density of misfit dislocations in the NWs has been observed, leading to a substantial reduction in their hole mobility to, i.e., only around 400 cm$^2$/Vs [32,34]. High crystalline-quality in-plane GeSi NWs, called GeSi hut wires (HWs), have been monolithically grown on Si(001) using molecular beam epitaxy (MBE) under *in-situ* annealing [35]. These HWs are fully strained and devoid of misfit dislocations. Based on this material, the first spin qubit in germanium was realized [9]. By combining top-down patterning and bottom-up self-assembly, we have further achieved site-controlled growth of GeSi HWs on Si [36,37]. However, these HWs can only be obtained at relatively high growth and annealing temperatures, typically in the range of 500-570 °C [35-38]. Such high-temperature processes lead to significant Si-Ge intermixing [39,40]. For the HWs on the planar Si substrate, their Ge content is lower than 70% [35]. On the patterned substrate, the Ge content is even lower due to enhanced intermixing caused by an increased availability of Si flow from the sidewalls in the patterned regions [41,42].

In this work, we report a method for low-temperature, site-controlled, epitaxial growth of in-plane pure Ge NWs with a high hole mobility by MBE. These Ge NWs are grown on prepatterned, strain-relaxed SiGe ridges on Si(001) at a temperature of 290 °C. The SiGe ridges can mitigate the lattice mismatch while simultaneously isolating Ge NWs from the surrounding Ge islands on the planar regions. Despite the significant reduction of the growth temperature, no defects are observed in the NWs by scanning transmission electron microscope (STEM) characterization. Electron energy-loss spectroscopy (EELS) analysis shows only Ge in the NWs. Electrical transport measurements reveal a hole mobility in the NWs of 7100 cm$^2$/Vs at 2 K, which is the highest value ever observed in 1D Ge systems. Furthermore, the cross-sections of the Ge NWs can be tuned by adjusting the size of the SiGe ridges and the amount of Ge deposited.

Figure 1a schematically illustrates the process of site-controlled growth of Ge NWs. We first fabricated ridge structures on a strain-relaxed Si$_{0.75}$Ge$_{0.25}$/Si(001) substrate using electron-beam lithography (EBL) and reactive ion etching (RIE). These ridges on the SiGe/Si substrate are oriented along a <100> direction and are about 100 nm in width and 80 nm in height. Here, we employ the SiGe/Si substrate instead of the conventional Si substrate, primarily for two reasons. First, the reduction in strain lowers the diffusion barrier for Ge atoms [43,44], promoting the formation of elongated NWs. Second, the lower strain effectively suppresses the generation of misfit dislocations and strain-induced intermixing [45]. After wet cleaning, the patterned substrate was loaded into the MBE chamber. A 60 nm Si$_{0.7}$Ge$_{0.3}$ buffer layer (BL) and a 4 nm Si spacer layer (SL) were then sequentially grown at temperature of 360-400 °C. Next, the Ge NWs were formed by depositing 3.5 nm Ge at 290 °C. Finally, a 3 nm Si capping layer (CL) was deposited at the same temperature to prevent oxidation of the Ge NWs. Notably, the thicknesses of both the compressively-strained Si$_{0.7}$Ge$_{0.3}$ BL and the tensile-strained Si SL are kept significantly below the Matthew-Blakeslee critical thickness of plastic strain relaxation [46,47]. Further details about the substrate patterning and the epitaxial growth are available in Methods (Supporting Information).

Figures 1b and 1c show bright-field STEM images of an as-grown Ge NW in cross-sectional and longitudinal views, where all the epi-layers are distinctly differentiated. As shown by the magenta dashed line in Fig. 1b, the interface between the SiGe BL and the Si SL exhibits a wiggly morphology. Note that from Figs. 3b-3e, we also observe wiggly Si/SiGe interfaces on the ridges with various widths. Such an uneven surface was formed after the SiGe growth on a ridge, which is detrimental to the subsequent growth of high-quality Ge NWs (see Fig. S3). However, by growing a few nanometers of Si at 400 °C, we are able to achieve a uniform (001)-faceted flat surface on the top. As indicated by white dashed lines in Fig. 1b and Figs. 3b-3e, this (001)-faceted surface is accompanied by two sidewalls with azimuth angles of approximately 37°, close to that of {304} facets. The Ge NW formed on this Si flat surface exhibits a {105}-faceted hut-shaped cross-section, similar to that of GeSi HW. The cross-sectional base size of the Ge NW follows the width of the Si flat surface $w_s$ (in the present case, $w_s \approx 55$ nm), while its height is approximately 7 nm. The Ge wetting layer on the sidewalls was measured to be approximately 1 nm thick and is disconnected from the surrounding planar regions. From the longitudinal cross-sectional STEM image of the Ge NW (Fig. 1c) and a top-view scanning electron microscopy (SEM) image of an individual NW (Fig. 1d), we find that our as-grown Ge NWs have a cross section which is uniform over their entire lengths except at their two ends.

The selective growth of Ge NWs on SiGe ridges, combined with the flexibility in substrate patterning, enables precise control over the position, distance, and length of the NWs. Figure 1e shows the morphology of a well-ordered NW array, demonstrating the effectiveness of this approach in achieving an array of Ge NWs with a high uniformity. Figure 1f illustrates a pair of closely spaced NWs, i.e., a double NW, obtained by directly reducing the spacing between two SiGe ridges, with an edge-to-edge distance of only about 80 nm. For Si and Ge, the [100] and [010] directions are crystallographically equivalent. Nanostructures can therefore be grown using a combination of these two directions. For example, as shown in Fig. 1g, a well-defined *L*-shaped Ge NW junction consisting of two NWs oriented perpendicularly to each other is obtained.

Figures 2a and 2b show high-angle annular dark-field (HAADF) STEM images of a Ge NW viewed along its length and cross-section, respectively. We see the high monocrystalline quality of all epi-layers. The right panel of Fig. 2a is the corresponding image obtained as inverse fast Fourier transform (IFFT) of the STEM image in the left panel, revealing perfect atomic columns with the (110) periodicity. These atomic columns are straight and coincide with that across different epitaxial layers, confirming the absence of misfit dislocations. This means the Ge NW and its surrounding Si layers remain strained and preserve the underlying SiGe lattice along its length ($y$-axis). The $xz$-plane strain distribution was further characterized by geometric phase analysis (GPA), as shown in the lower-left panel of Fig. 2b. Here, we took the $Si_{0.7}Ge_{0.3}$ BL as a reference to extract the relative strain components ($\varepsilon'$). From the strain map in $x$-direction, the $\varepsilon'_{xx}$ of the entire epi-layers is near zero, but it becomes positive and increases monotonically as it approaches the apex. Similar strain distribution has been observed in the self-assembled SiGe islands [48-50]. It indicates that the geometric strain relaxation occurs around the NW apex, while the remaining regions are still fully strained. From the $z$-direction strain map, we see that the Ge (Si) lattice, initially under in-plane biaxial compressive (tensile) strain, experiences out-of-plane expansion (contraction) via elastic relaxation.

The NWs' compositional distributions were studied by EELS measurements in STEM mode. Figure 2c shows EELS elemental maps and a line-cut. These results indicate that a pure Ge NW capped with pure Si has been successfully obtained through the strategy of lowering the growth temperature. In contrast, EELS data acquired for a GeSi HW which we have grown by high temperature annealing technique [35] on a planar Si(001) substrate suggests a peak Ge content of only 70% (see Fig. S8). Notably, the Si SL exhibits detectable Ge signals, which suggests that the Si-Ge intermixing is pronounced during the deposition of the Si SL at a temperature of 400 °C but not likely during the growth at a low temperature of 290 °C. These observations also align with the interface width characterization [48,49] of the successive epi-layers, as shown in the lower-right panel of Fig. 2.

The SiGe ridges may also offer us a possibility to control the orientation and cross-sectional shape of the grown Ge NWs. Therefore, we keep the growth conditions unchanged and carry out the growth by varying both the orientation and width of the ridges. We see that the morphology of the Ge NWs is strongly crystallographic orientation-dependent. High-quality, continuous NWs have only been achieved on the <100>-orientated ridges, while deviations from this direction have caused broken NWs (see Fig. S2). Then, we focus on the examination of the influence of the ridge width on the Ge NWs grown on <100>-orientated ridges. For this, the ridges with initial widths ($w_r$) of 60 nm, 70 nm, 80 nm, 90 nm, 100 nm and 110 nm were prepared on the SiGe/Si substrate. After the deposition of 4 nm Si, the corresponding Si flat surface widths ($w_s$) were measured to be approximately 15 nm, 25 nm, 35 nm, 45 nm, 55 nm and 65 nm, respectively. Figure 3a shows the as-grown morphologies of the Ge NWs, revealing that continuous NWs were formed with $w_r$ from 70 nm to 100 nm (corresponding to $w_s$ from 25 nm to 55 nm). However, with $w_r$ of 60 nm and 110 nm ($w_s$ of 15 nm and 65 nm), Ge islands and discontinuous NWs were observed, respectively. The cross-sections of these continuous NWs were characterized by STEM and the results are shown in Figs. 3b-3e. For the NW growth on the ridge with $w_r$ = 70 nm (Fig. 3b), an isosceles trapezoidal cross-section of the Ge NW was observed. Here, it is found that the Ge facets (indicated by yellow dotted lines) are parallel to the sidewalls of the Si SL and form continuous surfaces with the thin Ge wetting layer covering the Si sidewalls, which exhibit an inclination of approximately 37°. As $w_r$ increases, the {105} facets (green dotted lines in Figs. 3c-3e) appear at the top apex and extend laterally. When $w_r$ reaches to 100 nm, almost full {105} facets are formed on the Ge NW (green dotted lines in Fig. 3e).

The morphology of the Ge NWs has also been studied at different amounts of Ge deposition. Figs. 4a and 4b show top-view SEM images of the grown Ge NW samples obtained after deposition of different amounts of Ge on the ridges with $w_r$ = 90 and 70 nm, respectively. For $w_r$ = 90 nm, after 1.0 and 1.5 nm Ge deposition, Ge islands are observed, similar to the growth on a planar surface. As the amount of Ge deposition increases, these islands gradually grow and merge into wires. In contrast, for $w_r$ = 70 nm, we see the formation of continuous NWs after 1.5 nm Ge deposition. The cross-sectional STEM images of these continuous NWs are shown in the corresponding insets of Fig. 4. We see that, on both the 90 nm and the 70 nm ridges, {105}-faceted NWs are firstly formed and the NWs evolve towards the cross sections with steeper facets with a further increase of the deposited Ge amount.

Six field-effect transistor (FET) devices were fabricated from the {105}-faceted Ge NWs as shown in Fig. 2b and Fig. 3e. The transport properties of these NWs are characterized by electrical measurements in a $^4$He cryostat at temperatures of 2-100 K. The device fabrication started by selectively removing Ge islands around the NW using RIE to avoid current leakage, see the left panel of Fig. 5a for a Ge NW structure after the surrounding Ge islands were removed. Figure 5a (right panel) presents a false-colored SEM image of a typical Ge NW FET device and the measurement circuit setup. Details about the device fabrication electrical measurements are available in Methods (Supporting Information).

Figure 5b shows the two-terminal conductance ($G$) of a fabricated Ge NW FET device (device A) measured as a function of gate voltage ($V_g$) at an applied source-drain voltage $V_{sd}$ of 10 mV and a temperature of 2 K. It is seen that the conductance decreases with increasing $V_g$, showing the p-type conduction in the Ge NW channel as expected. The channel can be completely pinched off at a threshold voltage $V_{th}$ of ~1.6 V. At a sufficiently large negative $V_g$, the device total resistance is below 10 kΩ, indicating that good Ohmic contacts were formed. The contact resistance $R_c$ and the hole mobility $\mu$ can be extracted by fitting the measured $G(V_g)$ curve based on the equation [50-53]

$$G(V_g) = \left[R_c + \frac{L^2}{\mu C (V_{th} - V_g)}\right]^{-1}$$

. Here, $\mu$, $R_c$ and $V_{th}$ are the fitting parameters, and $C$ is the gate capacitance to the Ge NW channel, which was extracted based on a Poisson solver using finite element method and by taking the device geometrical parameters (see Fig. S9). Figure 5b shows the experimentally measured data (black curve) and the result of the fit (red curve). From the fit, we obtain a mobility of 7150 cm$^2$/Vs, which is one order of magnitude larger than the recently reported in-plane Ge NWs obtained by selective-area growth [32,34]. We note that previous NW FET studies [52,54] have shown that if considering the quantum confinement effect in the conduction channel, the $C$ was normally reduced by approximately 20%-30% from the value estimated classically based on a Poisson solver. Thus, our extracted mobility value of 7150 cm$^2$/Vs is most likely at its lower bound.

In addition, the elastic mean free path $l_e$ can be estimated based on the formula of $l_e = v_f \tau_e = \frac{\hbar}{e}\mu k_f$, where $v_f = \frac{\hbar k_f}{m^*}$ is the Fermi velocity, $\tau_e = \frac{m^* \mu}{e}$ is the elastic scattering time, $m^*$ is the effective hole mass, and $k_f$ is the Fermi wave vector. For the channel resistance in the range of 5 to 50 kΩ, $l_e$ is extracted to be in a range of 70 nm to 210 nm (see Supporting Information).

Figure 5c (left panel) shows the $G(V_g)$ curves of device A measured at temperatures ranging from 2 K to 100 K. With increasing temperature, the on-state $G$ decreases while the Ge NW channel threshold voltage $V_{th}$ increases. The Ge NW channel mobilities $\mu$ extracted from the measured $G(V_g)$ curves at different temperatures are shown in the right panel of Fig. 5c. Here it is seen that the channel $\mu$ exhibits two distinct temperature-dependent regions—a region with a slow decrease of $\mu$ with increasing temperature (2-20 K) and a region with a relatively fast decline of $\mu$ with increasing temperature (20-100 K). In the former region, $\mu$ is primarily determined by scattering from impurities at the dielectric/semiconductor interface and background impurities including lattice imperfections in the NWs and their surrounding Si layers. In the latter region, however, phonon scattering plays a dominant role. Nevertheless, it is seen that $\mu$ can still stay at a value of above 4000 cm$^2$/Vs at a temperature of up to 80 K.

Figure 5d shows a histogram of $\mu$ extracted from the fits to the measurements of $G(V_g)$ for all the six Ge NW FET devices at 4 K (see all the measured $G(V_g)$ data and the fitting curves in Fig. S10). Here, it is seen that the averaged value of $\mu$ at 4 K is 4800 cm$^2$/Vs with a distribution spanning from ~3000 cm$^2$/Vs to ~7000 cm$^2$/Vs. Smaller mobility values seen here are primarily due to inevitable RIE etching damage of the Ge NW channel during the process of removing surrounding Ge materials on the planar region of the substrate (see the cross-sectional TEM image of an FET channel shown in Fig. S9). The RIE etching process itself also introduces damage on the surfaces of devices, leading to an additional reduction in channel mobility. Here, we should stress that based on the aforementioned material analysis, i.e., the fact that no detectable Si signal within the NWs, absence of misfit dislocations, and sharp Si/Ge interfaces are observed in our as-grown Ge NW samples, we can firmly conclude that alloy scattering, defect scattering, and interface roughness scattering are all negligible when compared to the effects caused by the RIE etching damages. We are working on a new approach combining selective wet etching and surface passivation to optimize the process. In addition, given the fact that the NWs are capped with only a 3-nm Si, scattering from the proximal dielectric and its interface to top Si or SiO$_x$ layer could also play an important role in limiting channel mobility. Thus, increasing thickness of the Si or SiGe CL atop the NWs could be expected to substantially suppress the scattering and thereby further improve channel mobility.

In summary, we have successfully achieved self-assembled pure Ge NWs in MBE at low growth temperature by utilizing the prepatterned SiGe ridges on a strain-relaxed SiGe/Si(001) substrate. Leveraging the flexibility in substrate patterning, we have obtained a precise control of the position, spacing, length and cross-sectional dimension and shape of the NWs. Structural characterization and composition analysis have confirmed that the as-grown Ge NWs are of single crystalline quality and contain pure Ge. Electrical transport measurements of FET devices made from these Ge NWs reveal that the NWs can have a hole mobility of over 7000 cm$^2$/Vs in the temperature range of 2-20 K, which is a record-high value ever observed in Ge NW materials systems. It is strongly anticipated that these high-quality Ge NWs will boost the technological development of semiconductor-based quantum computing chips. Our innovative growth method may also be employed for assembling other novel semiconductor nanostructures for potential applications in various fields including nanoelectronics, optoelectronics and spintronics.

**Supporting Information**
Section 1. Methods.

Section 2. Growth of Ge nanowires.

Section 3. Structural characterization.

Section 4. Elemental analyses.

Section 5. Device structures and transport parameters.

**Corresponding Authors**

*E-mail: hqxu@pku.edu.cn.

*E-mail: jjzhang@iphy.ac.cn.

**Notes**

The authors declare no competing financial interest.

**Acknowledgements**

We thank Guilei Wang and Zhenzhen Kong from Institute of Microelectronics of the Chinese Academy of Sciences for providing the high-quality SiGe/Si(001) substrates. This work was supported by the Natural Science Foundation of China (NSFC) (Nos. 62225407, 92165207, 92165208, 11874071, 12304101, 12304207, 12304100) and the Innovation Program for Quantum Science and Technology (No. 2021ZD0302300).

**Figure 1**

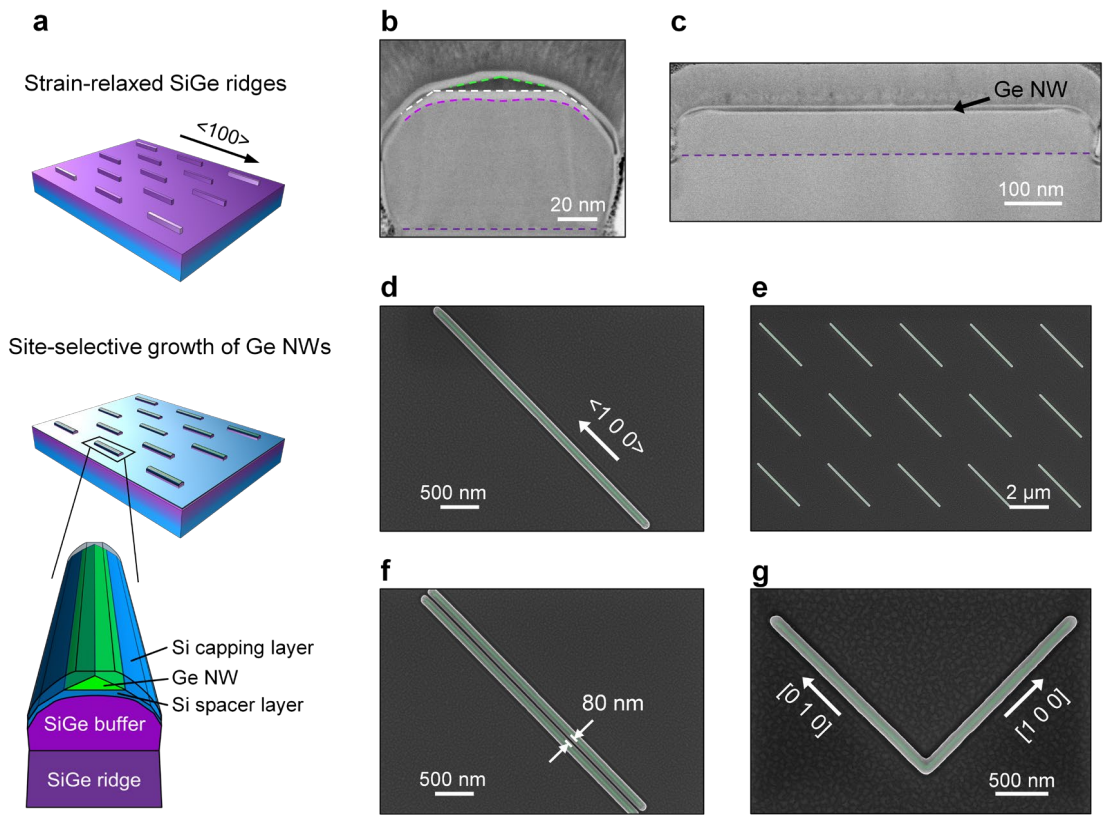

**Fig.1 Site-selective growth of Ge NWs on a ridge-patterned strain-relaxed SiGe/Si(001) substrate. a**, Schematics showing the strain-relaxed SiGe/Si(001) substrate with surface-patterned SiGe ridges (top panel), and the layer structures of a Ge NW sample achieved by area-selective growth on a SiGe ridge template (bottom panel). Here, the Si CL, Ge NW, Si SL, SiGe BL and initial SiGe ridge are marked in light blue, green, blue, magenta and purple, respectively. **b** and **c**, Bright-field STEM images of an as-grown NW in a cross-sectional and a side view, respectively. The interfaces of the epitaxial layers are marked with dashed lines. **d-g**, Top-view SEM images of an individual NW, a NWs array, a double NW, and an *L*-shaped NW junction formed by the intersection of [100]- and [010]-oriented NWs, respectively. The coloring has been applied to enhance the image contrast for Ge NWs in **d-g**.

**Figure 2**

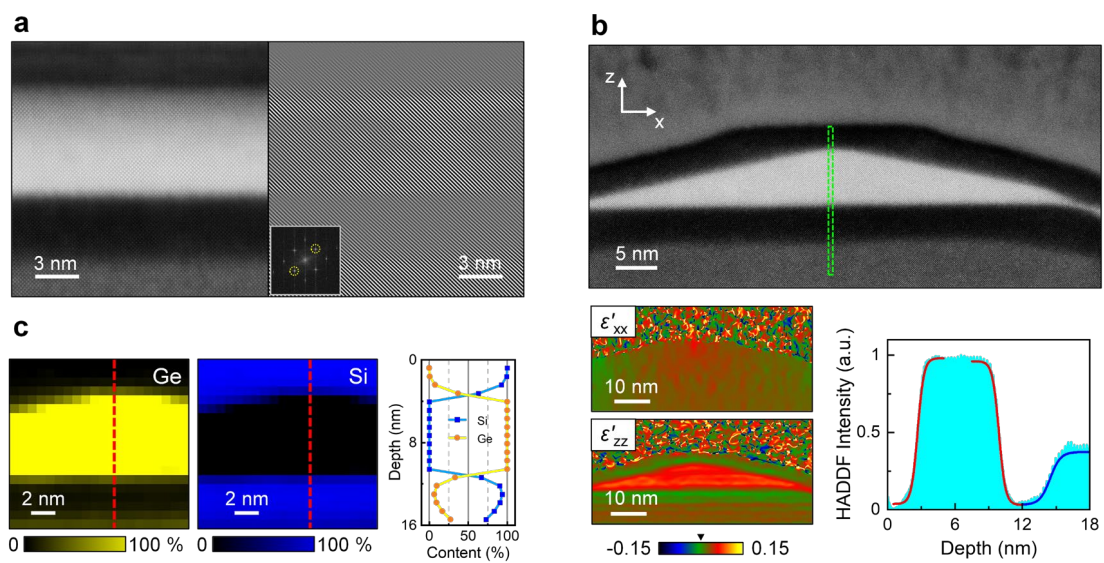

**Fig. 2 Structural characterization and compositional analysis of a Ge NW**. **a**, Atomic-resolution HAADF-STEM image of an as-grown Ge NW viewed along its length (left panel) and its corresponding IFFT image (right panel) obtained by extracting the (110) periodicity from the FFT spectrum. The inset shows the FFT spectrum of the atomic-resolution HAADF-STEM image, where the two yellow dotted circles mark the (110) periodicity spots in the spectrum. **b,** Cross-sectional atomic-resolution HAADF-STEM image of the Ge NW (top panel), its corresponding strain maps (lower-left panel) derived from GPA, and a HADDF intensity profile (lower-right panel) obtained from the region marked by green dotted frame in the top panel. The strain maps $\varepsilon'_{xx}$ and $\varepsilon'_{zz}$ were calculated using the lattice of SiGe BL as a reference. The solid curves superimposed on the HAADF intensity profile correspond to the sigmoid fits, which are employed for determining the interface width. From the fits, the interface width of both Si CL/Ge NW and Ge NW/Si SL are approximately 1.1 nm, while the interface of Si/Si$_{0.7}$Ge$_{0.3}$ BL is around 1.8 nm. **c**, EELS elemental maps obtained by extracting the Ge $L$-edge and the Si $K$-edge signals with a step length of 8 Å. The left to right panels show Ge map, Si map, and a line-cut showing the relative concentrations of Ge and Si along the red dashed lines in the two left panels.

**Figure 3**

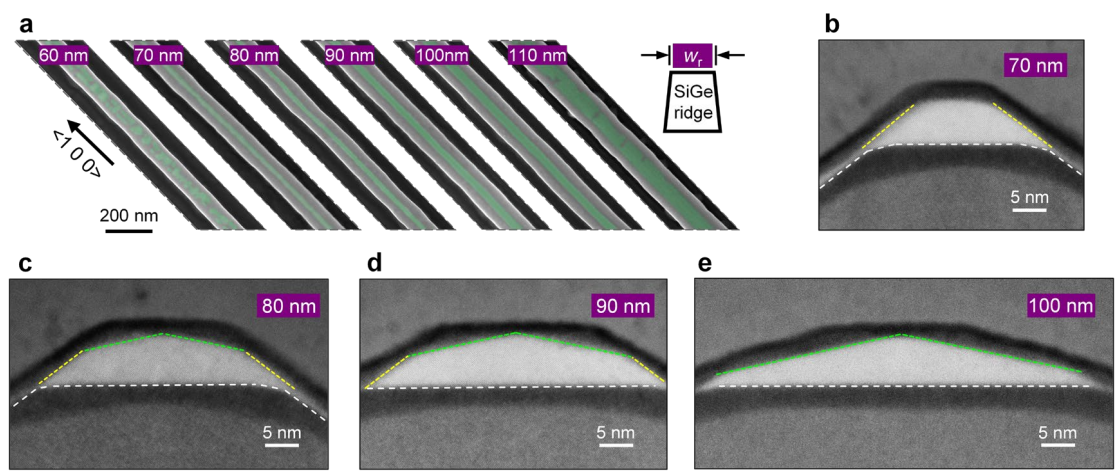

**Fig. 3 Tailoring the morphology of Ge NWs through varying the width of the SiGe ridges. a**, False-colored SEM images showing the morphologies of the Ge NWs grown on the <100>-orientated SiGe ridges with widths from 60 nm to 110 nm. Ge islands and discontinuous NWs are observed on the SiGe ridges with 60 nm and 110 nm in width, respectively, while continuous Ge NWs are formed on the SiGe ridges with 70-100 nm in width. **b-e**, Cross-sectional STEM images of the Ge NWs grown on the SiGe ridges with 70-100 nm in width, showing that the cross-sectional shape and size are tailorable. The interfaces are marked in dashed lines.

**Figure 4**

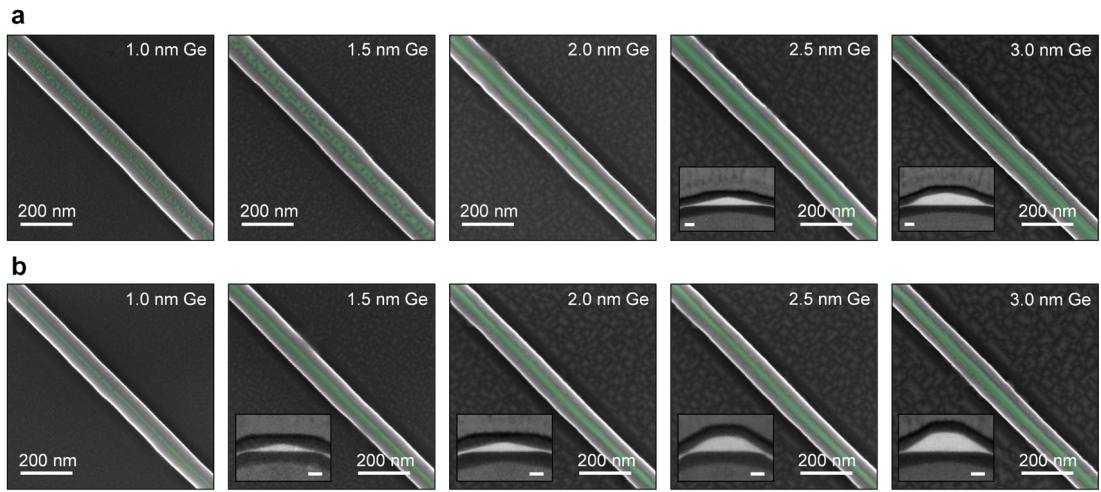

**Fig. 4 Morphology of Ge NWs tailored by the Ge deposition amount**. **a,b,** Top-view SEM images of the grown Ge NW samples obtained after deposition of different amounts of Ge on the ridges with $w_r$ = 90 nm and 70 nm, respectively. Here, the Ge deposition amount is incrementally increased from 1.0 nm to 3.0 nm in a step of 0.5 nm, and the SEM images are arranged from left to right accordingly. The insets show corresponding cross-sectional STEM images of the Ge NWs in the same panels, capped with 3 nm Si. These STEM images reveal the evolution of the cross-sectional shape of the Ge NWs with increasing Ge deposition. The scale bars in insets **a** and **b** are 5 nm.

**Figure 5**

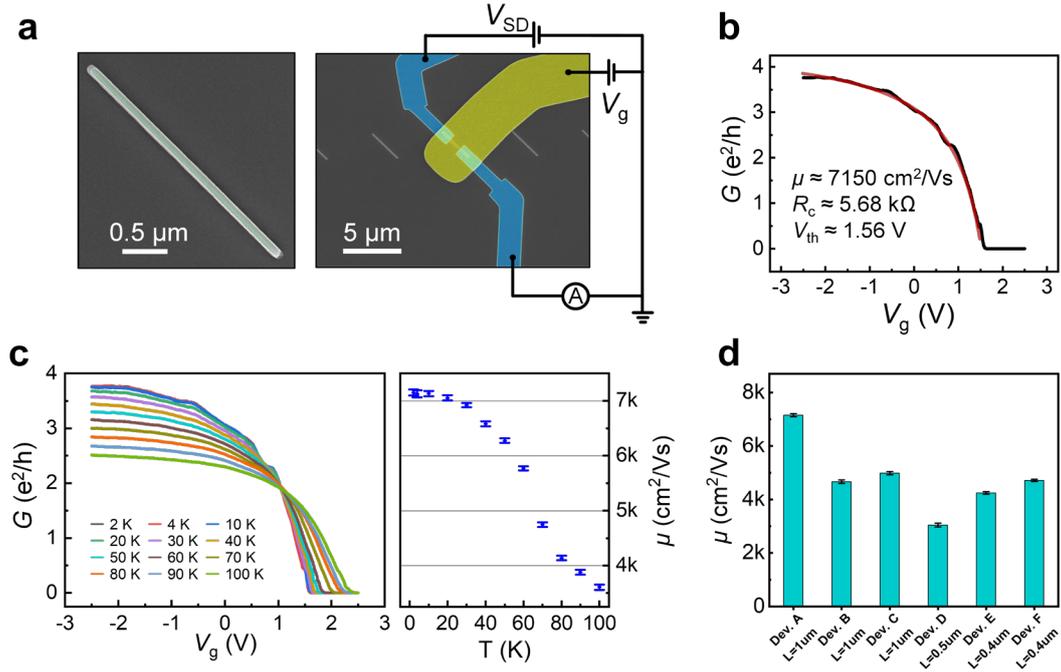

**Fig. 5 Electrical measurements of Ge NW FET devices**. **a**, SEM image of a Ge NW structure after selective removal of surrounding Ge islands on the flat region of the substrate (left panel), and false-colored SEM image of a typical Ge NW FET device made from a similar Ge NW structure as shown in the left panel and measurement circuit setup (right panel). The Pd contacts and the Ti/Au top-gate are colored in blue and yellow, respectively. **b**, Two-terminal conductance of a Ge NW FET device (device A) measured as a function of gate voltage $V_g$ at an applied source-drain voltage $V_{sd}$ of 10 mV and a temperature of 2 K. Mobility, contact resistance and threshold voltage are extracted from the fits (red curve) to the measurement data (black curve). **c**, Conductance of device A as a function of $V_g$ measured at an applied source-drain voltage $V_{sd}$ of 10 mV and in a temperature range of 2-100 K (left panel), and corresponding extracted mobilities at different temperatures (right panel). **d**, Histogram of mobilities extracted from the fits to the measurements of $G(V_g)$ for all the six fabricated Ge NW FET devices at 4 K. The average mobility is 4799.8 cm$^2$/Vs.

14